\documentclass[twocolumn]{article}
\usepackage{epsfig}
\begin{document}

\title{
{\huge \bf  
\center{Precision Cosmology? Not Just Yet\ldots}}\\
{\normalsize \bf 
Sarah~L.~Bridle, Ofer~Lahav, Jeremiah~P.~Ostriker, Paul~J.~Steinhardt}
\footnote{S. L. Bridle J. P. Ostriker and O. Lahav are at the 
Institute of Astronomy, Madingley Road, Cambridge, CB3 0HA, UK.
P.~J.~Steinhardt is in the Department of Physics, Princeton University, 
Princeton, New Jersey 08544, USA.}\\
\vspace{-1.5cm}
}
\author{}
\date{}
\maketitle

\noindent
The recent announcement by the WMAP satellite team of their 
landmark measurements of the cosmic microwave background (CMB)
anisotropy~\emph{(1-3)} has convincingly confirmed important
aspects of the current standard cosmological model.
The results show with high precision that space is flat (rather
than curved) and that most of the energy in the universe today is
``dark energy'', which is gravitationally self-repulsive and 
accelerates the expansion of the universe. The evidence is 
independent of supernovae results~\emph{(4,5)}.

The measurements strongly indicate that the amplitudes of
spatial variations in density and temperature that seeded the
formation of galaxies were roughly independent of length scale,
adiabatic (all forms of energy have the same spatial variation),
and followed a Gaussian distribution --- just as predicted by
the standard Big Bang inflationary model. WMAP heralds a new
age of precision cosmology with careful error analysis,
tightly constraining many key parameters~\emph{(6)}.
For example, the lifetime of the universe has been determined to
be $13,400 \pm 300$ million years~\emph{(6)}. Furthermore, 
WMAP's new measurement of the CMB polarization as a function of 
angular scale shows that the epoch of cosmic reionization ---
associated with the formation of the first stars --- had already
occurred when the universe was several hundred million years old.

At the same time we celebrate this triumph, it is important to 
recognize that important issues remain. For example, it is not
yet clear whether the spectrum of temperature fluctuations is 
truly consistent with inflation. The spectrum is roughly 
scale-invariant, but there are hints of peculiarities, and a 
key inflationary prediction --- the presence of gravitational 
wave effects --- has not yet been observed.

We also do not know whether dark energy is due to an unchanging,
uniform, and inert ``vacuum energy'' (also known as a cosmological
constant) or a dynamic cosmic field that changes with time and 
varies across space (known as quintessence). ``Dark matter'',
which is gravitationally self-attractive, also remains mysterious:
We do not yet know its nature, nor are we certain about its 
density or the amplitude of the initial ripples in its distribution.

Today's standard theoretical paradigm is the inflationary Big Bang
model. According to this picture, the universe began in a a state
of nearly infinite temperature and density and almost immediately
entered a phase of rapid, accelerated expansion (``inflation'').
This expansion smoothed out the distribution of energy, flattened
initial warp or curvature in space, and created tiny variations in
density. To transform these density variations into the gravitationally
collapsed, complex structures we see today, it is essential that
there be ``dark matter'' as well as ordinary (baryonic) matter.
Finally, we need dark energy to account for the measured total 
energy density and to explain the current cosmic acceleration.

\begin{figure}
\epsfig{ file=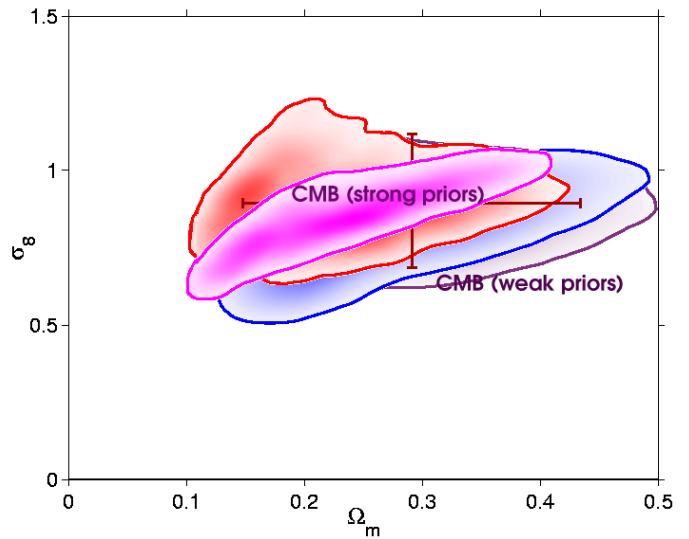, width=9cm}
\vspace{-0.2cm}
\caption{
\footnotesize
{\bf CMB constraints on $\Omega_{\rm m}$ and $\sigma_8$}. 
The pink contour corresponds to a ``strong prior'', 
which marginalizes over uncertainties in the Hubble constant,
baryon density, and spectral index of primordial fluctuations,
but assumes that other parameters are perfectly known,
including the optical depth to reionization, $\tau=0.17$.
The other contours are revised limits that include the
uncertainty in the equation of state of dark energy 
(blue, $-1<w<-0.7$); or $\tau$
[(red; in agreement with the WMAP alone constraint 
from~\emph{(6)}, shown by the red cross)]. 
The ``weak prior'' (purple) allows both these degrees of freedom.
All contours are 95\% confidence limits; shading corresponds
to the probability.
We used WMAP temperature and polarization data~\emph{(2,3,12)}
and small scale measurements from~\emph{(13-15)} and performed
the calculations with CosmoMC~\emph{(16)}.}
\vspace{-0.5cm}
\end{figure}

Some of the WMAP results -- the flatness of space, the near
scale-invariance, adiabaticity, and Gaussian distribution of the
density perturbations~\emph{(7)}, the density of baryons, the
age of the universe, and perhaps the early formation of the
first stars --- are based on WMAP alone and are consistent with
the standard model. Because the CMB is a direct image of the 
early universe and its interpretation entails simple, 
well-understood physical principles, these results are robust.

On the other hand, some important issues can only be addressed
by combining WMAP data with other cosmological measurements. 
These conclusions should be viewed more cautiously because the 
result depends sensitively on the choice of additional data.

For example, by combining data, a significant deviation from a
perfectly scale-invariant $(n=1)$ spectrum was found~\emph{(8)}.
According to the best-fit WMAP combined analysis~\emph{(8)}, 
$n$ runs from 1.1 on the largest scales to $<0.9$ on the smallest 
scales probed, a deviation that disagrees with the simplest
and most natural inflationary models~\emph{(9)}.
These results cast a pall over the inflationary paradigm at the
same time that many of its other important predictions are 
confirmed. However, it is important to note that WMAP data alone
are not inconsistent with the simplest inflationary models, the
fit being $n=0.99 \pm 0.04$. The inconsistency only arises when
the data are combined with two degree field galaxy redshift 
survey and quasar absorption line measurements.

Setting aside this issue, we next consider whether the WMAP 
data are uniquely consistent with the standard inflationary 
Big Bang picture. The answer is no, as the WMAP team has itself
indicated. There remains room for radical alternatives.
An example is the cyclic picture, in which the universe undergoes
a periodic sequence of cycles~\emph{(10)}: Expansion from a hot
big bang is followed by contraction in a ``big crunch'' and 
reemergence in a big bang, and the key events that shaped the
large-scale structure of the universe occur before the ``bang'',
a cycle ago. This model offers a very different view of cosmic
history, yet it fits all current observations (including the
new WMAP results) at least as well as the inflationary picture.

Even if the standard picture is proven to be correct, the
model is incomplete. The initial conditions that led to 
inflation and the identity of the ``inflaton'' field (the cosmic
field that causes inflationary expansion) remain unknown and the
nature of dark matter is unsettled.

An important uncertainty is the nature of the gravitationally
self-repulsive dark energy. Whether it consists of vacuum
energy or quintessence depends on $w$, the ratio of pressure
to energy density. The WMAP combined analysis concluded that
the best fit corresponds to vacuum energy. But their own 
closer examination of WMAP results at large angular scales
shows that the enhanced fluctuations expected for vacuum 
energy are missing. This discrepancy could be a first sign
that dark energy is quintessential~\emph{(11)}. The
combined analysis washes out the effect, but this could be
hiding a very important hint about the true nature of dark 
energy and confounding measurements of parameters.

WMAP has produced impressive constraints on many fundamental
cosmic parameters [see~\emph{(6)} for extensive tables], but
there remain open frontiers. Perhaps most important is the
uncertainty in the density of dark matter, $\Omega_{\rm m}$,
and the amplitude of density fluctuations labeled by $\sigma_8$,
as represented in the figures. These parameters are critical
because they describe the amount and distribution of the 
matter that clusters to form all of the structure in the 
universe.

The WMAP data alone only determine some combination of
$\Omega_{\rm m}$ and $\sigma_8$, represented by the narrow
pink contour in the first figure, if one takes into 
account the uncertainties in just three other cosmological
parameters but assumes the rest are perfectly known
(``strong prior''). Even with these over optimistic
assumptions, the range of $\Omega_{\rm m}$ and $\sigma_8$
is large. As we progressively relax the assumptions by
including empirical uncertainties, the range of degeneracy
balloons until we get to the more realistic purple contour
(the ``weak prior'').

\begin{figure}
\epsfig{ file=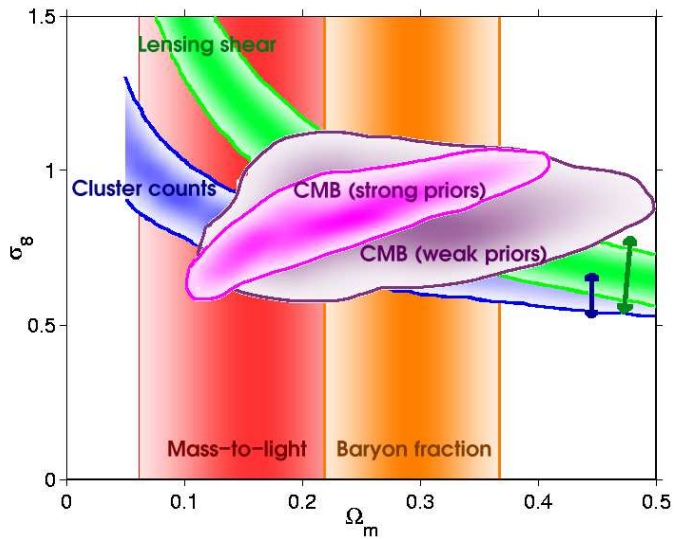, width=9cm}
\vspace{-0.2cm}
\caption{
\footnotesize
{\bf Comparison with other constraints.}
The ``strong prior'' (pink) and ``weak prior'' (purple)
WMAP constraints (from first figure) are compared
to other cosmological probes (95 \% confidence limits).
Red, mass-to-light ratio in galaxies and clusters of 
galaxies~\emph{(17)}. Orange, ratio of baryons to total 
matter content in clusters of galaxies~\emph{(18)}. 
Blue, observed numbers of galaxy clusters~\emph{(18)}
[blue arrow indicates scatter in results from different
groups, as reviewed in~\emph{(19)}].
Green, alignment of galaxy shapes in random directions in the sky
due to gravitational lensing [results from \emph{(20)}; 
spread of results reviewed in~\emph{(21)}, green arrow].
Other than the purple WMAP contour, all constraints are based
on the assumption that the Hubble constant, primordial 
spectral index, and dark energy parameter are well known.}
\vspace{-0.5cm}
\end{figure}

To reduce the uncertainty, the WMAP team has presented 
results obtained by combining with the two degree field
galaxy redshift survey and quasar absorption line measurements.
We agree with the mathematical conclusions obtained by the
WMAP team when combining these data sets. On the other hand,
it is worth considering the broader range of available data
shown in the second figure, which reveals three different 
directions of degeneracy: vertical strips from methods that
measure the matter density alone; constraints on the
combination $\sigma_8 \Omega_{\rm m}^{0.6}$, obtained by
measuring the number of galaxy clusters observed today
and the apparent distortion of galaxy images due to the
bending of light by dark matter; and a roughly orthogonal
constraint from the WMAP data. Here we see that the high 
likelihood regions (solid lines) do not all overlap well,
suggesting a problem with one or more of the measurements,
or their interpretation, or, more interestingly, that 
the underlying model may be wrong.

Perhaps adding only select measurements to WMAP data will
prove to be the correct strategy. On the other hand, given
the issues raised in the second figure, it may be that the
real uncertainty is much greater. We will have to wait for
forthcoming observations to substantially reduce the current
uncertainty.

Thus, even after the historic WMAP breakthrough, there 
remain unresolved issues, and so there is plenty of room
for surprises. Only 5 years ago, breakthroughs in 
technology and astronomical technique led to the
discovery that the expansion of the universe is accelerating.
The future holds promise of even greater technological 
advances that will uncover further cosmological surprises.

\medskip

\medskip

{\footnotesize
\begin{enumerate}
\vspace{-0.15cm}\item C.~Bennett~\emph{et al.} astro-ph/0302207
\vspace{-0.15cm}\item G.~Hinshaw~\emph{et al.} astro-ph/0302217
\vspace{-0.15cm}\item A.~Kogut~\emph{et al.} astro-ph/030221
\vspace{-0.15cm}\item S.~Perlmutter~\emph{et al.}, ApJ 517, 565 (1999)
\vspace{-0.15cm}\item A.G.~{Riess},~\emph{et al.}, AJ, 116, 1009 (1998)
\vspace{-0.15cm}\item D.~Spergel~\emph{et al.} astro-ph/0302209
\vspace{-0.15cm}\item E.~Komatsu~\emph{et al.} astro-ph/0302223
\vspace{-0.15cm}\item H.~Peiris~\emph{et al.} astro-ph/0302225
\vspace{-0.15cm}\item J.~Khoury, P.~Steinhardt, N.~Turok astro-ph/0302012
\vspace{-0.15cm}\item P.~Steinhardt, N.~Turok Science 296, 1436 (2002)
\vspace{-0.15cm}\item S.~DeDeo, R.~Caldwell, P.~Steinhardt astro-ph/0301284
\vspace{-0.15cm}\item L.~Verde~\emph{et al.} astro-ph/0302218
\vspace{-0.15cm}\item C.~Kuo~\emph{et al.} astro-ph/0212289
\vspace{-0.15cm}\item T.~Pearson~\emph{et al.} astro-ph/0205388
\vspace{-0.15cm}\item K.~Grainge~\emph{et al.} astro-ph/0212495
\vspace{-0.15cm}\item A.~Lewis, S.~Bridle PRD 66, 103511 (2002)
\vspace{-0.15cm}\item N.~Bahcall, R.~Cen, R.Dav{\' e}, J.Ostriker, Q.Yu ApJ 541, 1 (2000)
\vspace{-0.15cm}\item S.~Allen, R.~Schmidt, A.~Fabian, H.~Ebeling astro-ph/0208394
\vspace{-0.15cm}\item E.~Pierpaoli, S.~Borgani, D.~Scott, M.~White astro-ph/0210567
\vspace{-0.15cm}\item H.~Hoekstra, H.~Yee, M.~Gladders ApJ 577, 595 (2002)
\vspace{-0.15cm}\item M.~Jarvis~\emph{et al.} astro-ph/0210604
\vspace{-0.15cm}\item
We thank Antony Lewis for making the CosmoMC code publicly available
and for helping with the WMAP constraints.
Computations were performed at the UK National Cosmology 
Supercomputer Center funded by PPARC, HEFCE and Silicon Graphics 
/ Cray Research. 
Cluster counts constraints were kindly provided by S.~Allen.
We thank the Leverhulme Quantitative Cosmology group 
in Cambridge for helpful discussions.
\end{enumerate}
}
\end{document}